\documentclass[preprint]{aastex}

\newcommand{\vsi}{$v \sin i~$}
\newcommand{\ha}{H${\alpha}$}
\usepackage{epsfig}
\usepackage{graphicx}
\usepackage{natbib}
\usepackage{amsmath}
\shorttitle{The Circumstellar Disk of 48~Librae}
\shortauthors{Silaj et al.}

\begin{document}

\title{Investigating the Circumstellar Disk of the Be Shell Star 48~Librae}

\author{J.\ Silaj\altaffilmark{1}, C.\ E.\ Jones\altaffilmark{1},
  A.\ C.\ Carciofi\altaffilmark{2}, C.\ Escolano\altaffilmark{2},
  A.\ T.\ Okazaki\altaffilmark{3}, C.\ Tycner\altaffilmark{4},
  T.\ Rivinius\altaffilmark{5}, R.\ Klement\altaffilmark{5,6} \and
  D.\ Bednarski\altaffilmark{2}}

\altaffiltext{1}{Department of Physics and Astronomy, The University
of Western Ontario, 1151 Richmond Street, London, Ontario, N6A 3K7, Canada}

\altaffiltext{2}{Instituto de Astronomia, Geof\'{i}sica e Ci\^{e}ncias
  Atmosf\'{e}ricas, Universit\'{a}ria de S\~{a}o Paulo, Rua do
  Mat\~{a}o 1226, Cidade Universit\'{a}ria, 05508-900 S\~{a}o Paulo, SP Brazil}

\altaffiltext{3}{Faculty of Engineering, Hokkai-Gakuen University,
   Toyohira-ku, Sapporo 062-8605, Japan}

\altaffiltext{4}{Department of Physics, Central Michigan University,
  Mt.~Pleasant, MI 48859, USA}

\altaffiltext{5}{European Organisation for Astronomical Research in
  the Southern Hemisphere, Casilla 19001, Santiago 19, Chile}

\altaffiltext{6}{Astronomical Institute of Charles University, Charles
  University in Prague, V Hole\v sovi\v ck\'ach 2, 180 00 Prague 8}

\begin{abstract}
A global disk oscillation implemented in the viscous decretion disk
(VDD) model has been used to reproduce most of the observed properties
of the well known Be star $\zeta$~Tau.  48~Librae shares several
similarities with $\zeta$~Tau -- they are both early-type Be stars, they
display shell characteristics in their spectra, and they exhibit
cyclic $V/R$ variations -- but has some marked differences as well,
such as a much denser and more extended disk, a much longer $V/R$
cycle, and the absence of the so-called triple-peak features.  We aim
to reproduce the photometric, polarimetric, and spectroscopic
observables of 48~Librae with a self-consistent model, and to test the
global oscillation scenario for this target.  Our calculations are
carried out with the three-dimensional NLTE radiative transfer code
HDUST.  We employ a rotationally deformed, gravity-darkened central
star, surrounded by a disk whose unperturbed state is given by the VDD
model.  A two-dimensional global oscillation code is then used to calculate the
disk perturbation, and superimpose it on the unperturbed disk.  A very
good, self-consistent fit to the time-averaged properties of the disk
is obtained with the VDD.  The calculated perturbation has a period $P
= 12$ yr, which agrees with the observed period, and the behaviour of
the $V/R$ cycle is well reproduced by the perturbed model.  The
perturbed model improves the fit to the photometric data and
reproduces some features of the observed spectroscopic data.  Some
suggestions to improve the synthesized spectroscopy in a future work are given.
\end{abstract}

\keywords{circumstellar matter --- line: profiles --- radiative
  transfer --- stars: emission-line, Be --- stars: individual
  (48~Librae) --- stars: kinematics and dynamics}

\section{Introduction}

48~Librae (HD~142983, HR~5941) is a bright ($V \sim 5$) Be star that
began to display shell characteristics sometime between 1931 and 1935
\citep{far69}.  Its spectral features were already studied and
documented extensively by the mid-1960s (see, e.g., the review by
\citealt{und66}), but 48~Librae has continued to be the
subject of papers to the present day owing to the fact that
it possesses several characteristics that make it a particularly
intriguing object of study.

Be stars (sometimes referred to as Classical Be stars) are a
well studied class of objects consisting of near main-sequence,
rapidly rotating, B-type stars that have shown Balmer emission at some
epoch. The emission lines that characterize Be star spectra have been
conclusively shown to originate in a thin, equatorial disk consisting
of material that has decreted from the central star. Furthermore,
several recent studies have shown that these disks rotate in a
Keplerian fashion (see \citealt{mei074, del114, kra124, whe124}). 

The term ``shell star'' is reserved for a particular subset of Be
stars whose spectra display exceedingly narrow absorption cores
superimposed on broad emission lines.  They are understood as ordinary
Be stars viewed in an edge-on configuration; this provides a very
simple and natural explanation for their spectroscopic appearance, as
the narrow absorption cores arise due to the large optical depths that
occur when viewing the disk through its densest part.  In general,
shell stars are interesting objects to study because the uncertainty
in their inclination angle is largely eliminated.

In addition to being a shell star, 48~Librae displays a quasi-cyclical
asymmetry in its emission lines with the violet ($V$) and red ($R$)
peaks of the lines usually being of unequal height, and the peak
heights varying smoothly from configurations with $V > R$ to $V < R$.
Known as $V/R$ variations, this behaviour is common to about one-third
of all Be stars \citep{por03}, although 48~Lib possesses both a very
strong maximum flux and an unusually strong asymmetry in its
\ha\ profile.  The asymmetry can be so strong as to result in
misclassifications of the star as a supergiant (i.e. the line profiles
are misinterpreted as P~Cyg wind profiles).  An example of this can be
found in the SIMBAD database where the star is still listed as B8Ia/Ib
with a reference to \citet{1988mcts.book}.

48~Lib has long been known as a $V/R$ variable star, with the cycles
in its Balmer emission lines having a period of $\sim$10 years (see,
e.g., \citealt{men88}, and the many references therein). A recent
study of 48~Lib by \citet{ste12} suggests that the $V/R$ variations
were absent between 1995-1998, but that the cycle following that time
appears to have a period of $\sim$17 years, i.e. significantly longer
than previously observed.  Furthermore, they found the maximum
value of $V/R$ in the last cycle ($V/R \approx$ 2.2, which occurred in
2006) was greater than the maximum in the previous activity cycle.
For this study, we focus solely on the most recent cycle, and
therefore do not consider any data taken before 1995.

$V/R$ variations are thought to arise from a one-armed density wave
precessing in the disk \citep{oka97, car09, esc15}.  This necessarily
implies that the density distribution is non-axisymmetric in the disks
of stars exhibiting this phenomenon.  The differing $V$ and $R$ peak
heights seen in the line profiles are then explained very naturally by
the changing density configurations that are seen by a stationary
observer as the wave pattern precesses. The maximum value of $V/R$
observed in 48~Lib's last cycle is larger than what is usually
observed in Be stars, and suggests the density contrast in its disk
may be more extreme than typical.

48~Librae appears to be an isolated star, i.e. it has no known binary
companion.  Therefore, its disk, which we infer to be quite dense (on
average) from its large H$\alpha$ flux and strong IR excess, is very
likely to be quite extended.  While Be stars in binary systems may
have their disk truncated at the tidal radius by the secondary star
\citep{oka01,pan16}, there are no expected disk truncation mechanisms
present in 48~Lib.  This is considered to be somewhat of a rarity, as
a large fraction of Be stars are observed to exist as binaries or
multiple star systems \citep{san12}.

The viscous decretion disk (VDD) model, a description of the
circumstellar disk which was first introduced by \citet{lee91}, has
begun to emerge as the leading physical model of Be star circumstellar
disks.  In \citet{car09}, the VDD model was implemented in the
radiative transfer code HDUST to reproduce the spectroscopic,
polarimetric, photometric, and near-IR interferometric observations of
$\zeta$~Tau.  More recently, \citet{kle15} used the VDD model and the
HDUST code in an extensive study of $\beta$~CMi.  While those studies
demonstrated that the VDD model could successfully reproduce most of
the observables of each target, additional studies on varied targets
are required to determine if the VDD model describes Be stars
generally.

In this work, we use the VDD model in combination with the HDUST code,
described in \citet{car06,car08}, to investigate 48~Librae's
circumstellar disk.  This represents the first attempt at applying
this method to an isolated Be star with an extended disk, as both
$\zeta$~Tau and $\beta$~CMi are binary stars whose disks are
truncated.  Furthermore, it represents only the second attempt at
employing the model to reproduce cyclic $V/R$ variations. The work is
organized as follows: in Sect.~2, we describe our observations, and in
Sect.~3, we provide an outline of the computational codes employed,
followed by a discussion of our models of the central star and the
disk. In Sect.~4, we provide the results of our analysis, and in
Sect.~5, we give a summary and discussion.

\section{Observations}

\subsection{Photometry}

The VO Sed Analyser (VOSA; \citealt{bay08}) was employed to obtain
photometric values spanning from the UV to the far infrared.  We use
International Ultraviolet Explorer (IUE) HPDP photometry for the UV
portion of the SED, and Str\"{o}mgren $uvby$ \citep{hau98} and Johnson
$UBV$ \citep{mer94} photometry for the visible portion of the
SED. Tycho-2 \citep{hog00} and The Two Micron All Sky Survey (2MASS; \citealt{skr06}) photometry are used for the near infrared, and The
Wide-field Infrared Survey Explorer (WISE; \citealt{wri10}), The
Infrared Astronomical Satellite Mission (IRAS; \citealt{hel88}), and
AKARI \citep{mur07,ish10} data are used for the far-IR portion of the
SED.  It should be noted that the IRAS measurements at 60$\mu$m and
100$\mu$m are upper limits only \citep{wat87}.  \citet{tou13} report a
color excess $E(B-V) = 0.000 \pm 0.007$ for this star, and therefore
no dereddening was applied.  The photometric values and their
  respective errors are listed in Table~\ref{tab:phot}.

Clearly, the measurements used to create the SED are obtained
  at different epochs.  As Be stars are known to be variable objects,
  some variation in the measurements owing to the epoch in which they
  were taken is therefore expected. However, \textit{strong}
  photometric variations occur only for those Be stars whose H$\alpha$
  profiles undergo phase changes from emission to absorption, which
  are interpreted as disk-building (emission) and disk-loss
  (absorption) events.  48~Lib's H$\alpha$ profile has not been
  observed in absorption since it first became actively studied in the
  1930s, indicating a relatively stable disk that does not dissipate.
  In addition, the analysis of \citet{jon11} indicates 48~Lib's disk
  shows no clear trend of growth or loss.  Moreover, in our data
  spanning the most recent V/R cycle, we see no systematic changes in
  the observed H$\alpha$ equivalent width (EW), which again indicates
  that the disk is not building or dissipating.  For these reasons, we
  expect the photometry to remain fairly constant over different
  epochs, and anticipate only minor fluctuations in its values owing
  to the different configurations (due to the one-armed density
  oscillation) in which the disk may be viewed. This is confirmed by
  Figure~\ref{fig:SED}, where it can be seen that the maximal change
  in the model caused by the different disk configurations is slight,
  causing the predicted fluxes to vary more or less within the
  error bars associated with each measurement.

\subsection{Radio measurement}

Radio measurements exist for only a handful of Be stars, but are
extremely useful in determining the full extent of the circumstellar
disk.  This is because the IR and radio excesses in Be stars arise
mostly from free-free emission in the ionized gas.  As the free-free
opacity increases with wavelength, the flux at longer wavelengths
originates from progressively wider expanses of the disk
\citep{vie15}.  

48~Lib has been detected at 870 $\mu$m by the Atacama Pathfinder
EXperiment (APEX) \citep{gus06}, and the data were reduced with CRUSH
v. 2.31.  The value of the flux is given in Table~\ref{tab:phot} along
with the other photometric fluxes used in this work.  The application
of APEX to Be stars is described more fully in \citet{kle15}.

\subsection{Spectroscopy} \label{Spectroscopy}

High resolution H$\alpha$ profiles were collected from several sources
and analyzed in order to create the observed $V/R$ cycle. A total of
49 H$\alpha$ profiles were obtained on 30 nights between 2005 and 2015
with the fiber-fed \'{e}chelle spectrograph (also known as the Solar
Stellar Spectrograph) attached to the 1.1-meter John S. Hall telescope
at the Lowell Observatory, located near Flagstaff, Arizona. The
spectroscopic \'{e}chelle frames were reduced using a custom reduction
pipeline developed specifically for the instrument \citep{hal94}. 18
HEROS and 11 UVES measurements, obtained between 1995 and 2008, and a
single FLASH measurement, obtained in May 2000, were also employed.  A
detailed description of these observations and their reduction is
given in \citet{ste12}.  Finally, 54 H$\alpha$ line profiles (from 52
different nights) taken from the Be Star Spectra (BeSS) database
\citep{nei11} provided additional coverage of the $V/R$ cycle.  A
summary of the spectroscopic observations is given in
Table~\ref{tab:Ha}.

\subsection{Linear Polarimetry} \label{Polarimetry}

Polarimetric measurements of 48~Librae were obtained at the
Observat\'{o}rio Pico dos Dias (OPD), owned and operated by the
National Astrophysical Laboratory of Brazil (LNA).  A set of 14
measurements spanning from May 2009 to June 2015 that have $B, V$, and
$R$ or $B, V, R$, and $I$ values was compared to the polarization
predicted by our models. The polarimetric observations were performed
with the 0.6-m Boller \& Chivens telescope.  We used a CCD camera with
a polarimetric module described in \citet{m96}, consisting of a
rotating half-waveplate and a calcite prism. A typical observation
consists of 16 consecutive waveplate positions separated by 22\fdg5,
from which the Stokes parameters $Q$ and $U$ are obtained (see
\citealt{mag84} for details on the data reduction).

Initially, all polarimetric observations were corrected for the
interstellar polarization (ISP) using the same ISP value in
\citet{ste12}: $P_{\rm{max}} = (0.819 \pm 0.50)$\% at $\lambda =
5591$\AA\, and position angle PA = $(96 \pm 5)\degr$.  More recently,
\citet{dra14} have determined the ISP as $P_{\rm{max}} = 0.86$\% at
$\lambda = 5593$\AA\, with PA = $93\degr$.  We opted to also apply
this correction to our data and compare both sets of corrected
observations with the polarization predicted by our model.
Table~\ref{tab:pol} lists the full set of polarimetric values, with
$P_{\rm{1}}$ referring to the data corrected by the earlier ISP value,
and $P_{\rm{2}}$ referring to the data corrected by the newer ISP
determination.

\section{Theory and Models}

HDUST is a Monte-Carlo-based, fully three-dimensional (3D), non-local
thermodynamic equilibrium computer code.  It iteratively solves the
coupled problems of radiative transfer, radiative equilibrium, and
statistical equilibrium to obtain the temperature structure and level
populations for a gas of an arbitrary density and velocity
distribution.

The simulation proceeds by first emitting photons from a rotationally
deformed, gravity-darkened central star.  Each photon is tracked as it
travels through the circumstellar disk until the point that it
eventually escapes.  During its travel, the photon interacts (usually
many times) with the gas via scattering, or absorption and reemission.
These three processes depend on the gas opacity and emissivity, and we
include both continuum processes and spectral lines in the
determination of these gas properties. In the case of scattering, the
photon changes direction, Doppler shifts, and becomes partially
polarized.  In the case of absorption, the photon is not destroyed,
but is reemitted locally with a new frequency and direction determined
by the local emissivity of the gas. Since photons are never destroyed,
radiative equilibrium is automatically enforced, and flux is conserved
exactly.  For a full discussion of the HDUST code, the interested
reader is referred to \citet{car06,car08}.

\subsection{The Central Star}

Consulting the literature, it was found that the spectral type
classifications of 48~Lib are numerous and varied.  For instance, it
has been classified as B3V by \citet{und53}, B5IIIp shell He-n by
\citet{les68}, B6p shell by \citet{mol72}, B4III by \citet{jas80},
B3IV:e-shell by \citet{sle82}, and as B3 shell by \citet{jas92}, just
to name a few.  Accurate spectral classification of Be stars is known
to be particularly difficult due to the presence of emission lines and
high rotational velocities which may fill in, distort, or even
entirely obliterate key lines normally used in the determination.
Hence, to commence our modelling, we initially treated the spectral
type of the central star as a free parameter and considered central
stars with parameters consistent with B3-B6 stars of various
luminosity classes.

In order to translate a spectral type into physical measurements for
the star, we consulted \citet{har88} to first obtain the mass
($M_\star$) that best corresponds to each numeric subclass. From Table
4 of the aforementioned work, the masses of a B3, B4, B5, and B6 star
are 6.07, 5.12, 4.36, and 3.80~$M_\odot$, respectively.  It should be
noted that these masses are somewhat smaller than those provided by
\citet{cox00}. We elected to employ the masses determined by
\citet{har88} as this study represents a very thorough investigation
of stellar parameters that focused expressly on B-type stars.
Furthermore, this source was also used to fix a mass value for
$\zeta$~Tau in \citet{car09}, and we aim to follow the modelling
procedure of that paper as closely as possible in order to facilitate
comparisons between the two works.

For each of the four masses given above, we then employed the Geneva
stellar models of \citet{geo13}, which are a set of evolutionary
models specifically designed for rapidly rotating stars, to determine
the polar radius ($R_{\rm{p}}$) and luminosity ($L_\star$) that
corresponds to each subtype.  Once a mass has been supplied to the
stellar evolution code, only the metallicity, age, and fractional
rotation rate of the star are required in order to calculate all other
parameters: the polar radius, equatorial radius, luminosity,
temperature, fractional surface abundances, etc., which are all
given as a function of time.  In all cases, we assumed a solar
metallicity ($Z_\odot = 0.014$) for our central star.  As the \vsi
measurements of 48~Lib are remarkably homogeneous, with most authors
finding a value of 400~km\,s$^{-1}$ (see, e.g., \citealt{sle75, sle82,
  bal95, bro97, cha01, abt02}), we employed this value to set the
fractional rotation rate, {$v_{\rm{frac}} =
  v_{\rm{eq}}/v_{\rm{crit}}$, with $v_{\rm{crit}} =
  \sqrt{2GM_\star/3R_{\rm{p}}}$. Assuming that 400~km\,s$^{-1}$ is a
  good approximation to the linear velocity at the stellar
  equator\footnote{This value may be an slight underestimation, as
    gravity-darkening may cause a narrowing of spectral lines
    \citep{cra05}.  This, in turn, would mean the true value of
    $\omega_{\rm{frac}}$ is slightly greater than 0.95, although our
    models would still necessarily adopt the value of 0.95 as values
    greater than this are not provided by the Geneva stellar models.},
  $v_{\rm{eq}}$, (i.e. assuming that $i$ is sufficiently close to
  $90\degr$ that sin~$i\approx 1$), we find that
  $v_{\rm{eq}}/v_{\rm{crit}} \approx 0.80$ for the models with masses
  of 6.07 and 5.12~$M_\odot$.  For the models of masses 4.36 and
  3.80~$M_\odot$, the fractional velocity is $\approx 0.83$ and 0.85,
  respectively.  (It should be noted that the approximation arises not
  only because the $v_{\rm{eq}}$ value is approximate, but because the
  actual values of $M_\star$ and $R_{\rm{p}}$ for each model will vary
  with time.  The value of $v_{\rm{frac}}$ is estimated from values
  approximately representing the middle main sequence of the star's
  lifetime.)  The Geneva models require that the rotation velocity be
  specified in terms of $\omega_{\rm{frac}} =
  \Omega$/$\Omega_{\rm{crit}}$ (where $\Omega_{\rm{crit}} =
  \sqrt{8GM_\star/27R^3_{\rm{p}}}$).  Converting our linear velocities
  to angular ones, we find $v_{\rm{frac}} \approx 0.80 \rightarrow
  \omega_{\rm{frac}} \approx 0.95$ and $v_{\rm{frac}} \approx 0.83$ or
  $0.85 \rightarrow \omega_{\rm{frac}} \approx 0.96$.  However, the
  Geneva models are available only for a set of discrete
  $\omega_{\rm{frac}}$ values, with $\omega_{\rm{frac}} = 0.95$ being
  the maximum allowed value, and so this value was adopted for all
  models.

In the HDUST code, the star can be defined by its mass, rotation rate
(given in terms of $W = v_{\rm{rot}}/v_{\rm{orb}}$, see Sect. 2.3.1 of
\citealt{riv13}), polar radius, and luminosity combined with just one
other parameter -- the gravity-darkening exponent -- to describe the
star in a completely self-consistent way.  We use a model of
fast-rotating stars in the Roche approximation, and the spectrum of
each latitude bin (described by a pair of $T_{\rm{eff}}$ and
$\rm{log}~g$) is obtained from an interpolation of the Kurucz
models. For the description of the photospheric model see
\citet{car108}.

To select the best value for the gravity-darkening
exponent ($\beta_{\rm{GD}}$), we consulted the work of \citet{esp12},
who showed that the standard von Zeipel law, $T_{\rm{eff}} \propto
g^{\beta_{\rm{GD}}}_{\rm{eff}}$ with $\beta_{\rm{GD}} = 0.25$
\citep{von24}, is only valid in the case of spherical stars, and that
very rapidly rotating, isolated stars represent the strongest
deviation from this law.  Interpolating from their Table 1, we obtain
$\beta_{\rm{GD}} = 0.17$ for 48~Lib.

\subsection{Disk Structure}

The disk is assumed to exist in the equatorial plane of the star,
which is perpendicular to its rotation axis. As per the standard
convention, the angle $i$ denotes the inclination to the observer's
line of sight, with $i = 0\degr$ corresponding to a pole-on
orientation, and $i = 90\degr$ corresponding to an edge-on viewing of
the circumstellar disk. Since the disk is not axisymmetric, an
additional angle $\varphi$ (the azimuthal angle) is employed to denote
the phase at which the disk is being viewed.

\subsubsection{Unperturbed Disk State}
To quantify the densities required to reproduce each of the peak
heights in the observed H$\alpha$ profile, we employed a symmetric
disk whose density structure represents the steady state VDD, and
varied the values of initial disk density at the stellar surface.  It
was assumed that the surface density of the disk decreases as an $n =
2.0$ power law with increasing radial distance $r$ from the central
star, i.e. $\Sigma = \Sigma_0 (r/R_\star)^{-n}$, where $\Sigma_0$ is
an assumed initial surface density at $r = R_\star$.  It should be
noted that $R_\star$ corresponds to the equatorial radius of the star,
which is quite different from the polar radius ($R_{\rm{p}}$) owing to
the rapid rotation of the star. The vertical structure of the disk is
described by a Gaussian density distribution
\begin{equation}
 \rho = \frac{\Sigma(r)}{\sqrt{2\pi \cdot H(r)}} e^{-(1/2)(z/H(r))^2}
\end{equation}
with a radial power law for the disk scale height,
\begin{equation}
H = H_0(r/R_\star)^{\beta}.
\end{equation}
The radial exponent $\beta$, which is also known as the disk flaring
parameter, was set to 3/2 in all simulations, for which the
temperature is constant throughout the disk.  $H_0$, the disk scale
height, is given by
\begin{equation}
H_0 = \frac{a}{v_{\rm{orb}}} \cdot R_\star,
\end{equation}
where $v_{\rm{orb}}$ is the orbital velocity and 
\begin{equation}
 a=\sqrt{\frac{k_BT_0}{\mu m_H}}
\end{equation}
is the isothermal sound speed.

In the last equation, $k_B$ is the Boltzmann constant, $\mu$ is the
mean molecular weight, and $m_H$ is the mass of a hydrogen atom.
$T_0$ is the isothermal disk temperature that sets the vertical scale
height.  Typically, $T_0$ is set to a fractional value of the
effective temperature of the central star, $T_{\rm{eff}}$, and we
adopt $T_0 = 0.7\,T_{\rm{eff}}$ \citep{car06}.  We note that using this
prescription for the surface density and scale height is equivalent to
setting $n = 3.5$ in the expression $\rho(r,0) =
\rho_0(r/R_\star)^{-n}$, which is another way that the density
structure of Be star disks is typically described.

Finally, the disk is assumed to be composed entirely of hydrogen, and
its velocity structure is assumed to be Keplerian rotation.  Disks
produced by this formalism assume an axisymmetric density
distribution.

\subsubsection{Perturbed Disk State}
In order to model the $V/R$ variations, we employ the two-dimensional
(2D) global oscillation model of \citet{oka97} and \citet{pap92}.
This code calculates perturbed surface densities by imposing an $m$=1
perturbation on the unperturbed state, which is discussed above,
resulting in a non-axisymmetric disk that possesses both overdense and
underdense regions.  The theory and assumptions of this code are
described at length in \citet{car09}, but we recall here some of the
main inputs. The values that were adopted for each of these inputs are
summarized in Table~\ref{tab:48params}.

In calculating the gravitational potential, the global oscillation
code considers the quadrupole contribution due to the rotational
deformation of the rapidly rotating central star.  Thus, the potential
is given by
\begin{equation}
\psi \simeq -\frac{GM}{r} \bigg\{1 + k_2
\bigg(\frac{\Omega_{\star}}{\Omega_{\rm{crit}}}\bigg)^2
\bigg(\frac{r}{R_{\star}}\bigg)^{-2}\bigg\}
\end{equation}
where $\Omega_{\star}$ is the angular rotation speed of the star,
$\Omega_{\rm{crit}} = \frac{2}{3}(R_{\rm{p}})^{-1}V_{\rm{crit}}$, and
$k_2$ is the apsidal motion constant.

The equation of motion in the radial direction is used to derive
$\Omega(r)$, the radial distribution of the rotational angular
velocity:
\begin{equation}
\Omega(r) \simeq -\bigg(\frac{GM}{r^3}\bigg)^{1/2} \bigg[1 + k_2
  \bigg(\frac{\Omega_{\star}}{\Omega_{\rm{crit}}}\bigg)^2
  \bigg(\frac{r}{R_{\star}}\bigg)^{-2}
  + \frac{\rm{d\, ln}\,
    \Sigma}{\rm{d\, ln}\,
    r}\bigg(\frac{H}{r}\bigg)^2-\eta\bigg(\frac{r}{R_{\star}}\bigg)^{\epsilon}\bigg)\bigg]^{1/2},
\end{equation}
under the approximation $z^2/r^2 \ll 1$.  Note that this expression
includes a radiative force term
\begin{equation}
F_{\rm{rad}} = \frac{GM}{r^2}\eta\bigg(\frac{r}{R_{\star}}\bigg)^{\epsilon},
\end{equation}
where $\eta$ and $\epsilon$ are parameters that describe the force due
to an ensemble of optically thin lines \citep{che94}.

As shown in \citet{car09}, we use the equation
\begin{equation}
\Sigma_0 =
\frac{\dot{M}}{3\pi\alpha{c_s}^2}\bigg(\frac{GM}{R_{\star}}\bigg)^{1/2}\bigg[\bigg(\frac{R_0}{R_{\star}}\bigg)^{1/2}
  - 1 \bigg],
\end{equation}
where $\alpha$ is the viscosity parameter of \citet{sha73}, $c_s$ is
the sound speed, and $R_0$ is a parameter related to the physical size
of the disk, and $\dot{M}$ is the mass-loss rate of the central
star.

\section{Results}

\subsection{The SED}

We have collected the observed spectral energy distribution (SED) of
48~Lib from the UV to the radio regime (Figure~\ref{fig:SED}).  The
flux measured at any given wavelength, of course, represents a
combination of the flux from the central star and flux from the disk.
Our modelling has revealed, however, that the UV portion of the SED
specifically is only marginally affected by the adopted model disk
density and instead depends mostly on the parameters of the central
star.  Once the central star parameters have been set, the observed IR
excess can be matched by increasing the initial density $\rho_0$ of
the model disk until agreement is obtained. Due to the large
computational expense required to run perturbed models, we first fit
the IR portion of the SED with an unperturbed disk to determine the
\textit{average} disk density that reproduces the IR excess before
applying a perturbation to the model.

The observed parallax of 48~Lib is $p = 6.97 \pm$ 0.24 mas
\citep{van07}, which translates to a distance $d = 143\substack{+5
  \\ -10}$\,pc.  Using the standard flux-luminosity relation and
fitting the models of different spectral types to the SED, we find
that the observed UV fluxes are best reproduced by a B3V central star
of mass $M_\star = 6.07M_\odot, R_{\rm{p}} = 3.12R_\odot$ and $L_\star
= 1100L_\odot$. We note that the parameters adopted for the central
star form a consistent set, but are not necessarily unique in their
ability to reproduce the observed properties.  Altering any one of
these values by a small amount (e.g. 10\%), the remaining parameters
could likely be adjusted so as to still reproduce the observations.

Stars of spectral types later than B3 were simply not luminous enough
to match the observed flux levels, even accounting for the increased
luminosity as the star evolves, while a more evolved B3 star (i.e. of
luminosity class IV or III) was too luminous to match the
observations. In Figure~\ref{fig:uv}, the UV portion of some of the
tested model spectral types are compared with the IUE measurements. In
all cases, the model spectra are computed assuming the central star is
surrounded by a dense disk, and that the system is oriented at
85\degr. Main-sequence stars of spectral type B2($M_\star =
8.62M_\odot$) to B6 are shown in solid lines.  In the case of the B6V
model, we also show the effect of changing the distance to the star
within $\pm 1\sigma$ of the Hipparcos parallax.  The lower dash-dotted
line represents $p - \sigma$ and the upper dash-dotted line represents $p
+ \sigma$.  Each model experiences the same spread in its flux by
adjusting the distance in this manner, and thus it can be seen that
errors in the distance do not imply a significant error in the stellar
parameters.  The best fit to the data is clearly given by the B3V
model SED.  A more evolved B3 star (i.e. a B3IV star) is also shown
(black dashed line) on the plot, but it is obviously too luminous to
reproduce the observations.  The B4IV model (blue dashed line) appears
to match some of the IUE data points, but it is noticeably
underluminous in the far UV portion of the SED.  Note also that the
overall shape of the model B4IV SED is somewhat flattened and does not
reflect the observed shape.  Models of B5IV and B6IV stars (not shown)
suffered this same discrepancy, and were systematically underluminous
to reproduce the observations.

The observed IR fluxes are well matched by a disk whose initial
density is $\rho_0 = 8.8 \times 10^{-11}$~g\,cm$^{-3}$ at the stellar
surface (Figure~\ref{fig:SED}) and oriented at 85\degr\,
(Figure~\ref{fig:48libpol}).  We again note that this density represents only
the average (unperturbed) initial disk density.  While our first
estimate of the average initial disk density was obtained by fitting
the IR portion of the SED, we further constrained this value by
consulting the H$\alpha$ profile shown in Figure~\ref{fig:ha48}.  The
final value of the average initial disk density was obtained in an
efficient manner by modelling the individual peak heights of this
profile and then averaging the two values.  Notably, the density
derived from the H$\alpha$ line was in good agreement with the value
obtained by fitting the IR portion of the SED.  The H$\alpha$ line,
however, was more sensitive to small changes in density, thus helping
to better constrain the model.  The H$\alpha$ observations and the
modelling procedure are discussed further in Section~ \ref{sec:Ha}.

A B3V central star with a disk whose average density was obtained by
the methods described above matches the UV (IUE data) and visible
portion (Str\"{o}mgren $ubvy$, Johnson $UBV$, and Tycho-2 data) of the
SED exceedingly well. The Balmer jump at $\lambda = 3646$\AA~ is well
matched by a model oriented at 85\degr, which is further supported by
the polarization measurements (see next section).  The only data point
not well matched in this portion of the SED is the Str\"{o}mgren $u$
flux at $\lambda = 3500$\AA.  However, we note that this measurement
has a large associated error, and that the upper limit of the
measurement is well matched by our model.  Furthermore, the Johnson
$U$ measurement, which is at a similar wavelength but has a much
smaller error, is also well fit by our model, suggesting that the flux
near this wavelength is indeed closer to the upper limit of the
Str\"{o}mgren $u$ measurement.

The infrared portion of the SED merits some discussion.  In the near
IR (Tycho-2 data), the assumed average disk density again fits the
observations remarkably well.  In the far IR (WISE, AKARI, and IRAS
data) however, there is a small discrepancy between the unperturbed
model and the observed data points.  We note that the IRAS data points
at 60 and 100 microns are only upper limits to the fluxes at those
wavelengths \citep{wat87} and should not be given much weight in
determining the best disk density.  The IRAS measurements at 12 and 25
microns, in addition to the WISE and AKARI data points, however, all
show an improved fit when the perturbed model is applied to the SED.

48~Librae is one of only a few Be stars to have a radio (APEX data)
measurement. The APEX measurement is of particular interest because it
is well matched by both the unperturbed and perturbed models, and it
falls more or less in line with the other data points in the
SED. \citet{vie15} discussed the effects of truncation on the IR SED.
According to their Fig.~11, disk truncation alters the slope of the IR
SED at wavelengths for which the radius of the pseudo-photosphere,
which is the disk region vertically optically thick in the continuum,
is of the order of the truncation radius.  The fact that our model
reproduces well the SED up to radio wavelengths suggests that
48~Librae is an isolated Be star, i.e. it does not experience
disk truncation by a companion.

\subsection{Polarization}

As shown in \citet{car09}, the continuum linear polarization levels
provide a strong constraint on the inclination angle $i$ of the
star+disk system. The left panel of Figure~\ref{fig:48libpol} shows 14
polarization observations, all obtained within the most recent $V/R$
cycle, at $\lambda$ = 4450, 5510, 6580, and 8060\AA. The observations
in this plot have been corrected by two different ISP values, as
discussed in Section~\ref{Polarimetry}.  The considerable spread in
the observed polarization level at each wavelength is to be expected,
as polarization levels are extremely sensitive to the changing density
configurations in the disk that occur with $V/R$ variations
\citep{hal13} and to short-term variations in the disk feeding rate
\citep{hau14}. As done in our fitting of the SED, we analyze the
polarization by attempting to match it with an unperturbed model.} The
predicted polarization level of the unperturbed model that fits the
SED (Figure~\ref{fig:SED}) is shown for four different inclination
angles: 75\degr, 80\degr, 85\degr, and 90\degr. At $\lambda =
4450$\AA~, the $P_{\rm{1}}$ data points cluster around the 85\degr\,
model or slightly lower, while the $P_{\rm{2}}$ data points centre on
the 85\degr\, model.  We give some preference to using the $P$ values
at $\lambda = 4450$\AA~, as it can clearly be seen in the figure that
the model changes most noticeably at this wavelength, whereas the
differences in models become consistently smaller at longer
wavelengths.  As shown in the right panel of
Figure~\ref{fig:48libpol}, the average of each reduction of the
polarization is fairly well fit by the model at 85\degr, although at
$\lambda = 4450$\AA~ the average of $P_{\rm{1}}$ is better fit by a
slightly higher inclination angle (87\degr), and at $\lambda =
5510$\AA, the average of $P_{\rm{2}}$ is better fit by a slightly
lower inclination angle (82\degr). The averages of $P_{\rm{1}}$ and
$P_{\rm{2}}$ are not very well fit at $\lambda = 6580$\AA, but it can
be seen from the left panel that adopting a lower inclination angle
does not improve the fit to these two points; in fact, the model
created at 75\degr\, has a lower $P$ value at this wavelength than the
model created at 80\degr.  All data points are fairly well fit by any
of the models at $\lambda = 8060$\AA, but we note that the models
experience a fair bit of degeneracy at that point, as the differing
inclination angles cause almost no discernable difference in the
models.  We therefore give preference to fitting the data at $\lambda
= 4450$ and $5510$\AA.  By bracketing the lowest $P_{\rm{1}}$ value at
$\lambda = 4450$\AA~ and the highest $P_{\rm{2}}$ value at $\lambda =
5510$\AA, we determine the inclination angle to be approximately $85
\pm 3$\degr.

\subsection{The H$\alpha$ Profile}
\label{sec:Ha}

The H$\alpha$ profile shown in Figure~\ref{fig:ha48} was obtained
2005-04-01. The violet peak has $F/F_c \approx 5.2$ while the red peak
has $F/F_c \approx 2.5$, yielding a $V/R$ value of $\approx 2.1$,
which is very close to the maximum value of $V/R$ reported in 2006.
This indicates that, at the time of the observation, 48~Lib's disk is
oriented such that the overdense region is moving toward
the observer almost in its entirely, while the underdense region is
receding from the observer.

To quantify the densities required to produce each peak height, we
model them separately.  The violet peak height is best represented by
a model with an initial density $\rho_0 = 1.1 \times
10^{-10}$~g\,cm$^{-3}$, and the red peak height is reproduced by a
model with an initial density $\rho_0 = 6.5 \times
10^{-11}$g\,~cm$^{-3}$. Averaging the two densities, we obtain $\rho_0
= 8.8 \times 10^{-11}$g\,~cm$^{-3}$. We note that an axisymmetric
model of initial density $\rho_0 = 1.1 \times 10^{-10}$~g\,cm$^{-3}$
reproduces the large peak at some times, but would, on average,
produce profiles too large to match the observation, while the initial
density of $\rho_0 = 6.5 \times 10^{-11}$g\,~cm$^{-3}$ produces
profiles too weak on average.  This can be seen by comparing
Figure~\ref{fig:ha48}, where the lower density profile has peaks only
at about $F/F_c \approx 2.7$ and the higher density profile has peaks
at $F/F_c \approx 5.5$, with the centre panel of
Figure~\ref{fig:phases} where $V = R$ and $F/F_c$ is about 3.6 or 3.7.
However, these values of initial density can be interpreted as a good
first order approximation to the upper and lower limits to the range
of densities in our model that produces a reasonable facsimile of the
observations.

\subsection{$V/R$ Variations}

The model disk structure predicted by the 2D global oscillation code
is shown in Figure~\ref{fig:pert}.  This model is obtained by
superimposing an $m=1$ oscillation on the unperturbed disk of initial
density $\rho_0 = 8.8\times10^{-11}$g\,cm$^{-3}$.  The values of the
input parameters $k_2$, $\eta$, $\epsilon$, $\alpha$, $\dot{M}$ are
listed in Table~\ref{tab:48params}.  The value of $k_2$ is taken from
\citet{pap92}, but is, in theory, a free parameter.  The adopted
values of $\eta$ and $\epsilon$ are those that were used the
successful modelling of $\zeta$~Tau in \citet{car09}.  The value of
$\alpha$ is a free parameter that may vary between 0 and 1, with lower
$\alpha$ values usually corresponding to more confined modes.  The
parameter $\alpha$ ultimately governs the mass decretion rate, $\dot{M}$.
We adopted the value of $\alpha = 0.76$ as it was shown in
\citet{esc15} that higher $\alpha$ values tend to better reproduce Be star
observables, and it yields a mass decretion rate that is consistent with
what is expected to maintain a dense and extended disk such as the one
surrounding 48~Lib.  The oscillation produced by these five parameters
has a period of 12.13 years, which is in good agreement with the
average observed $V/R$ cycle length of $\sim$10 years.

48~Librae's most recent $V/R$ cycle is shown in
Figure~\ref{fig:vrcycle}. One full cycle is defined as the time
between the $V/R$ minimums. The first such minimum occurs shortly
after MJD 50000, and the second occurs shortly after MJD 560000, for a
total cycle length of approximately 6000 days, or $\sim$16.5 years.
The $V/R$ cycle corresponding to one full precession of the model disk
is also shown in Figure~\ref{fig:vrcycle}.  Given the complexity of
the model, and the large number of input parameters, the agreement
between the observed and model cycles is quite remarkable; the overall
behaviour of the observed cycle is matched quite well in a qualitative
sense. There are, however, some manifest discrepancies between the
model and the observations.  Firstly, the amplitude of the $V/R$ cycle
is not matched by the model.  Our model displays the same minimum
$V/R$ ($\approx 0.5$) as the observed one, and its maximum $V/R$
($\approx 2.1$) is close to the observed one ($\approx 2.4$), but it
does not attain the same value. Furthermore, the width of the main
peak is too broad in the model, in the sense that the model predicts a
rise in $V/R$ slightly before it is observed, and the decline of $V/R$
also happens more slowly in the model than what is observed.

In an attempt to determine if the amplitude of the $V/R$ could be
matched by a greater disk density, we computed a model with the
initial disk density increased by 30\%.  This resulted in only a
marginally larger maximum $V/R$ value ($\approx 2.15$), but produced a
$V/R$ cycle with drastically different behaviour that no longer
matched the general features of the observation.

A sample of model H$\alpha$ profiles are compared with observed
profiles (obtained at the Lowell observatory) in
Figure~\ref{fig:phases}.  The left panel shows an observation obtained
2007-03-26 and shows the near maximum $V > R$ configuration in the
disk.  This configuration is fairly well matched by our model,
although the red peak is somewhat overestimated by the model.  The
centre panel shows an observation obtained 2009-07-10, when $V \approx
R$.  The peak intensity is quite well matched; however, the central
absorption is not nearly strong enough in our model.  Finally, the
right panel shows an observation obtained 2012-06-22 when $V > R$, and
as in the leftmost panel, the weaker peak is again overestimated by
the model.

In each of the three panels in Figure~\ref{fig:phases}, one clear
discrepancy in the model is the inability to fit the observed line
wings.  The very broad wings that are seen in the observations are
thought to arise mainly from non-coherent scattering, which is not
included in our model, and thus this discrepancy is to be expected.

An additional discrepancy common to all of the panels in
Figure~\ref{fig:phases} is that the central absorption of the model is
too small to match the observations.  This feature most likely arises
from the fact that our model assumes a pure hydrogen composition for
the disk.  As shown in \citet{sig07}, including metals in the disk
model can lead to a predicted temperature distribution that differs
significantly from the one that is predicted for a pure hydrogen disk.
Since our model does not account for the the cooling effect that
metals provide, the temperature in the outer parts of the model disk
is most likely overestimated.  This, in turn, leads to an
overestimation of the flux in the outer portion of the model disk, and
artificially fills in the central absorption.

We further compared the model and observed H$\alpha$
line profiles by considering their EWs.  To do so, we employed the
carefully measured EWs of the Lowell Observatory H$\alpha$
observations.  As mentioned in Section~\ref{Spectroscopy}, these
observations were taken on 30 nights between 2005 and 2015, and
samples most of the most recent $V/R$ cycle.  It was found that the EW
of the H$\alpha$ line varied between $\sim-26$ \AA\, and $\sim-18$
\AA\, during that time, with the average of all the measurements being
-21.8 \AA.  Averaging the EWs of the H$\alpha$ line profiles obtained
from one full model $V/R$ cycle, we obtain the value -21.3 \AA.  This
indicates that the density obtained by modelling each of the peaks
individually is approximately correct, and reproduces the strength of
the H$\alpha$ emission.

\section{Summary and Discussion}

By examining and modelling several key observables simultaneously, we
obtained a self-consistent description of 48~Librae and its dense,
extended circumstellar disk.  Our final stellar and disk parameters
are summarized in Table~\ref{tab:48params}.

One of the first goals of this work was to model the SED of 48~Librae
in order to conclusively determine the star's spectral type;
accurately determining the central star's parameters is crucial to
modelling the entire system, as the photoionizing radiation provided
by the central star affects all other aspects of the model. The SED,
however, cannot be modelled in its entirety without considering the
disk. For this reason, we opted to achieve a first estimate of the
average disk density by separately modelling the two peaks of one
representative H$\alpha$ profile with simple, axisymmetric models, and
then averaging the two initial densities that were obtained.  This
proved to be an efficient method of determining the average initial
density of the disk.  The SED was well matched by the axisymmetric
model created with this average initial density, and applying the
perturbation actually improved the SED fit.  From the SED fitting, we
conclude that the central star is of spectral type B3V, and it is
surrounded by a disk of average initial density $\rho_0 = 8.8
\times10^{-11}$ g\,cm$^{-3}$.

The modelling of the SED also gives an indication of the inclination
of the system.  For a disk of a given density distribution, the
adopted inclination will affect the size of the Balmer jump at
$\lambda = 3646$\AA\, \citep{bjo05}.  The SED was best fit by
models oriented at 85\degr, and this $i$ was then confirmed by
examining the polarization levels.  This finding supports the
hypothesis that shell stars are Be stars viewed in a nearly edge-on
configuration.

The values of the densities obtained in this work are of some
interest, as they are all considered to be quite high.  This suggests
48~Lib's disk is unusually dense, which is congruent with its very
large flux in H$\alpha$ and its strong IR excess.  The density
contrast in the two regions responsible for the violet and red
H$\alpha$ emission peaks is also remarkable.  Our analysis implies
that the overdense region is more than 1.7 times as dense as the
underdense region.

An additional interesting result of this work is the observed radio
flux of 48~Librae, which implies a large disk size.  Using the
pseudo-photosphere model of \citet{vie15} we estimate that at
870$\mu$m the size of the disk pseudo-photosphere of our best fit
model is about 35$R_\star$, which implies that the disk cannot be
smaller than this value.  In any case, this finding is not
inconsistent with 48~Librae being an isolated Be star.

The 2D global oscillation code predicts a perturbed disk with a period
of $\sim$12 years, which is in good agreement with previously observed
$V/R$ cycle lengths observed for 48~Librae, although the most recent
$V/R$ cycle appears to have had a period of $\sim$16 years.
Furthermore, the general behaviour of the $V/R$ cycle is well matched
by the model, and the EWs of the H$\alpha$ lines produced by the model
are also in excellent agreement with the observed ones.  Some aspects
of the model $V/R$ variations do require some improvement, however.
Firstly, the amplitude of the $V/R$ cycle could not be matched by our
model, and testing a model with increased initial density did not
rectify this.  In fact, the model with increased initial density
displayed entirely different behaviour in its model $V/R$ cycle, which
indicates that small changes in the model can have large and
unanticipated effects.  Secondly, the main peak of the $V/R$ cycle is
too broad in our model.  Interestingly, this effect was also observed
by \citet{car09} when they applied the 2D oscillation model to
$\zeta$~Tau.  It seems this phenomenon may be a feature of the 2D
model, as \citet{esc15} were able to properly reproduce the peak of
the $V/R$ cycle by employing a 2.5D oscillation.  It is our intention
to test this hypothesis in a future work.

One further aspect of the observations that was not particularly
well matched by the model is the shapes of the H$\alpha$ line profiles
shown in Figure~\ref{fig:phases}.  As mentioned above, we plan to
employ a 2.5D (or, ideally, a fully 3D) oscillation in a future work,
which should increase the realism of our models and perhaps resolve
this issue as well. A factor that may prove more important in
improving the appearance of the H$\alpha$ profiles, however, is the
use of a version of HDUST that employs a realistic solar chemical
composition.  This version of HDUST is currently being developed.
Finally, we also hope to improve and refine all aspects of our model
by employing H$\alpha$ and near-IR interferometry to place tighter
constraints on the disk geometry and density configuration in a future
work.  This first attempt at applying a global oscillation scenario to
48~Lib has proven quite successful at reproducing most of the
observables, and executing the improvements suggested above should
lead to a much better understanding of the stellar and circumstellar
components of this unusual Be star.

\acknowledgments

We would like to thank an anonymous referee for their helpful comments
and suggestions that strengthened and improved the paper.  This
research has made use of the SIMBAD database, operated at CDS,
Strasbourg, France, NASA's Astrophysics Data System, the BeSS
database, operated at LESIA, Observatoire de Meudon, France:
http://basebe.obspm.fr, and VOSA, developed under the Spanish Virtual
Observatory project supported from the Spanish MICINN through grant
AyA2011-24052. This publication makes use of data products from the
Two Micron All Sky Survey, which is a joint project of the University
of Massachusetts and the Infrared Processing and Analysis
Center/California Institute of Technology, funded by the National
Aeronautics and Space Administration and the National Science
Foundation. This publication also makes use of data products from the
Wide-field Infrared Survey Explorer, which is a joint project of the
University of California, Los Angeles, and the Jet Propulsion
Laboratory/California Institute of Technology, funded by the National
Aeronautics and Space Administration. This work is based partly on
observations collected at the European Southern Observatory, Chile
(Prop. Nos. 54.D-0114, 56.D-0381, 266.D-5655, 081.C-0475).  We thank
the Lowell Observatory for the telescope time used to obtain the
H$\alpha$ line spectra presented in this work, and C.~T. acknowledges,
with thanks, grant support from the Central Michigan
University. A.~C.~C. acknowledges support from CNPq (grant
307594/2015-7) and FAPESP (grant 2015/17967-7 and 2011/51720-8).
C.~E.~J. and J.~S. thank the FAPESP/Western Researcher Award, Proposal
ID 0000029280, for travel funding to facilitate this project with
their colleagues in Brazil. The research of R.~K. was supported by
grant project number 1808214 of the Charles University Grant Agency
(GA UK) and by the research program MSM0021620860 (M\v{S}MT \v{C}R).
This work has made use of SHARCNET, the Shared Hierarchical Academic
Research Computing Network in Canada, and the computing facilities of
the Laboratory of Astroinformatics (IAG/USP, NAT/Unicsul), whose
purchase was made possible by the Brazilian agency FAPESP (grant
2009/54006-4) and the INCT-A. This research was supported in part by
NSERC, the Natural Sciences and Engineering Research Council of
Canada.

\begin{figure}
\centering
\includegraphics[scale=.5]{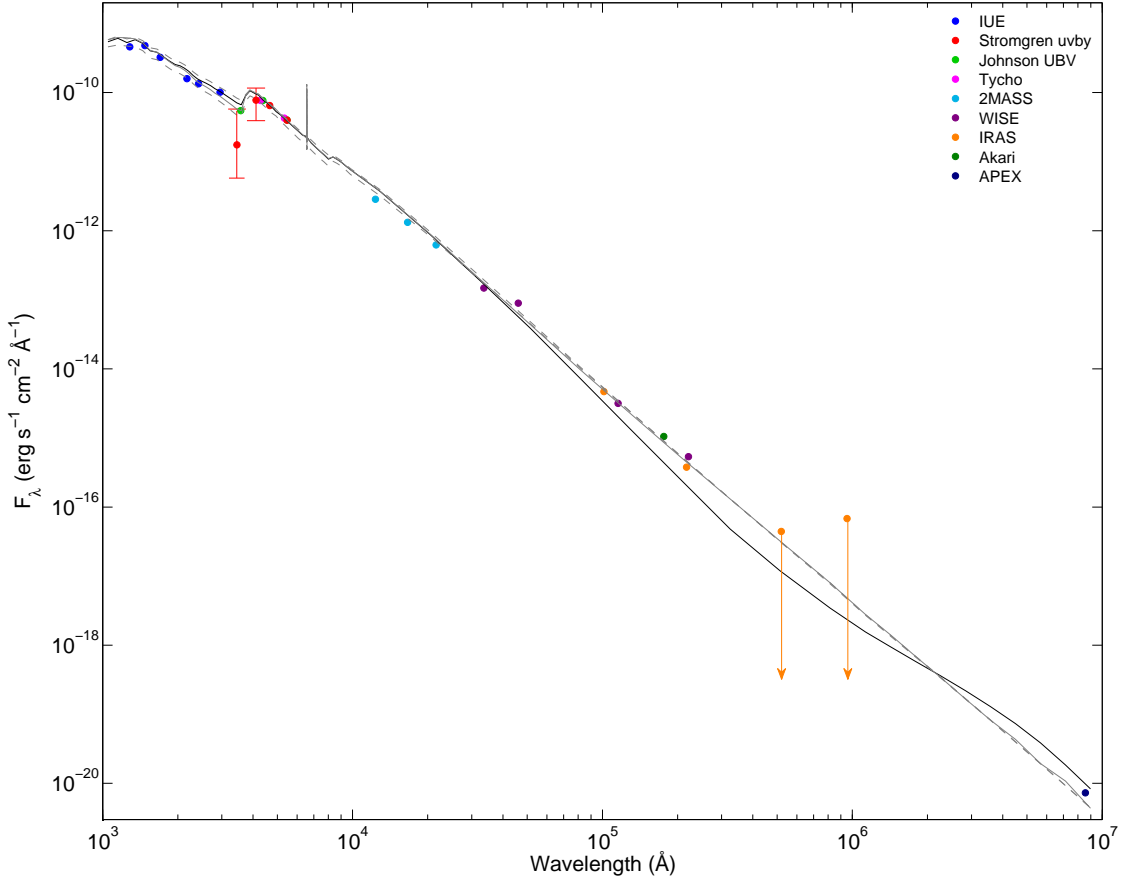}
\caption[SED of 48~Librae]{SED of 48~Librae. Filled circles represent
  measured fluxes obtained with various instruments, as indicated in
  the legend.  The size of the points is equal to or slightly greater
  than the errors in the measurements, except in the two Str\"{o}grem
  points where error bars have been explicitly added.  The two IRAS
  data points at 60$\mu$m and 100$\mu$m are considered upper limits
  only.  The solid black line represents the unperturbed model created
  for a B3V star with an average initial disk density $\rho_0 = 8.8
  \times 10^{-11}$~g\,cm$^{-3}$ oriented at 85\degr.  The solid grey
  line represents the perturbed model created with the same model star, initial
  density, and orientation, in a $V = R$ configuration.  The dashed grey
  lines represent the extrema that the $V/R$ variations cause in the
  model SED, with the upper line representing a $V < R$ configuration
  and the lower line representing a $V > R$
  configuration.\label{fig:SED}}
\end{figure}

\begin{figure}
\centering
\includegraphics[scale=.6]{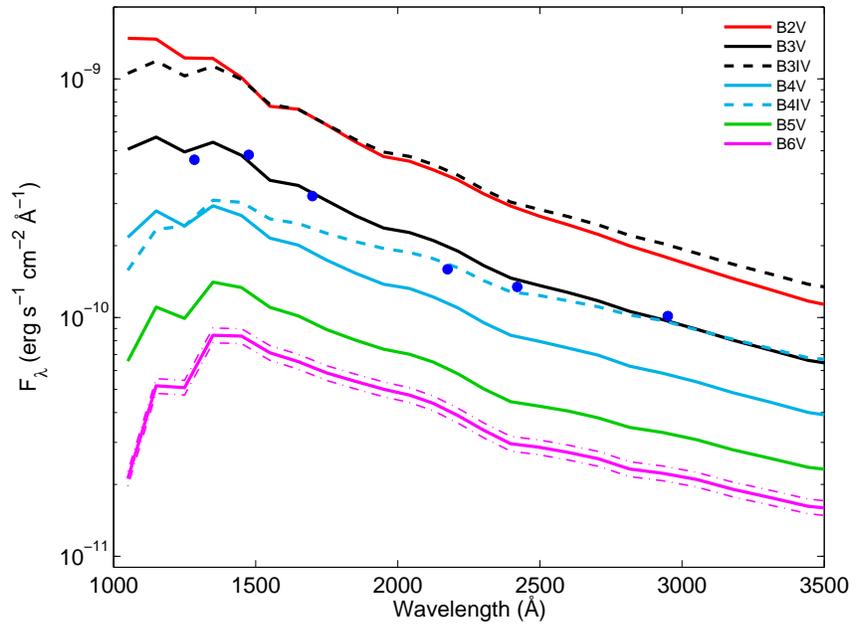}
\caption{IUE measurements and tested model spectral types for
  48~Librae.  IUE measurements are represented by filled blue circles,
  and the synthetic SEDs for various spectral types are shown by solid
  or dashed lines as indicated in the legend.  The dash-dot lines
  above and below the B6V model represent the SED computed at the
  distance given by $p + \sigma$ and $p - \sigma$,
  respectively.\label{fig:uv}}
\end{figure}

\begin{figure*}
\includegraphics*[width=3.45in]{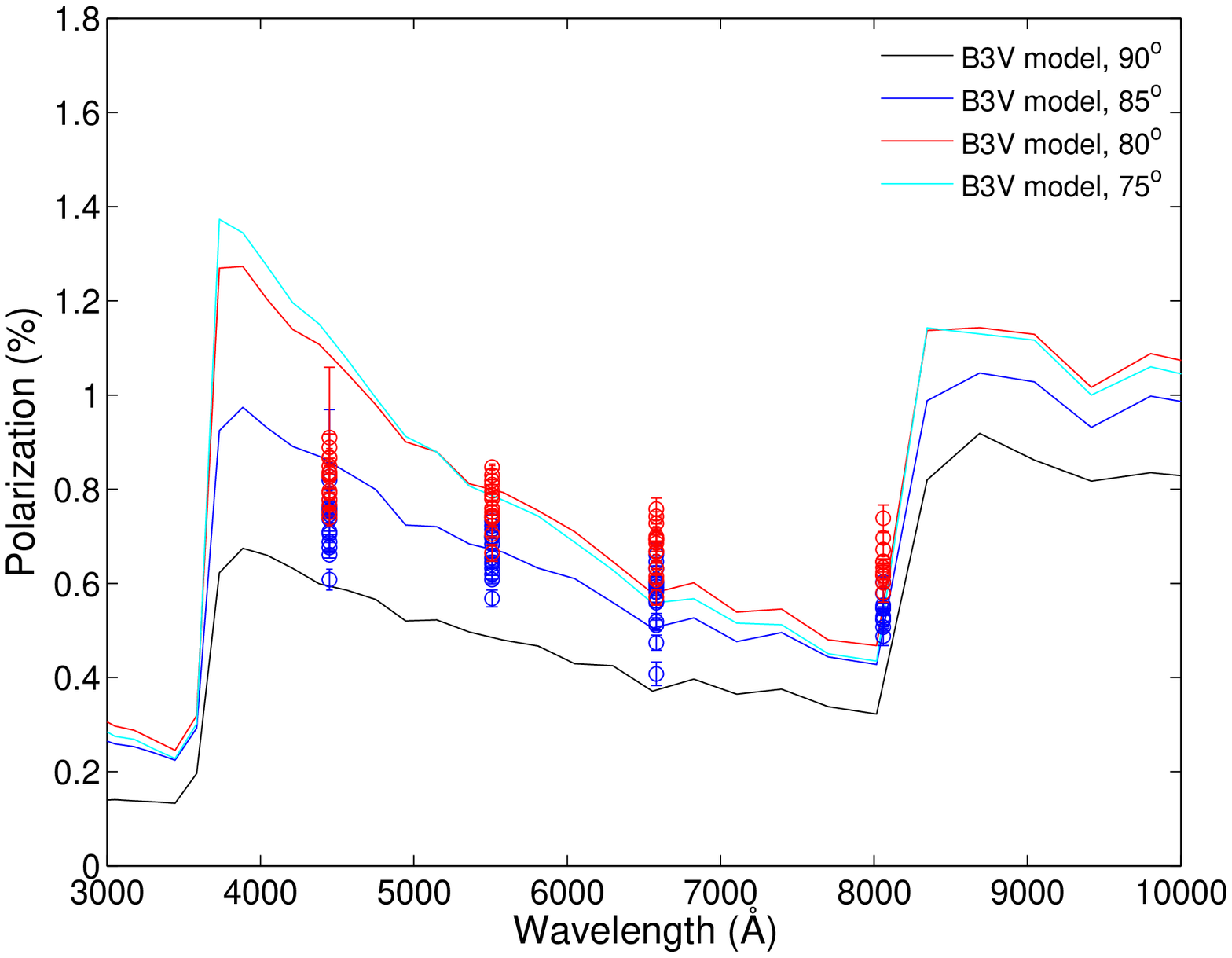}
\includegraphics*[width=3.45in]{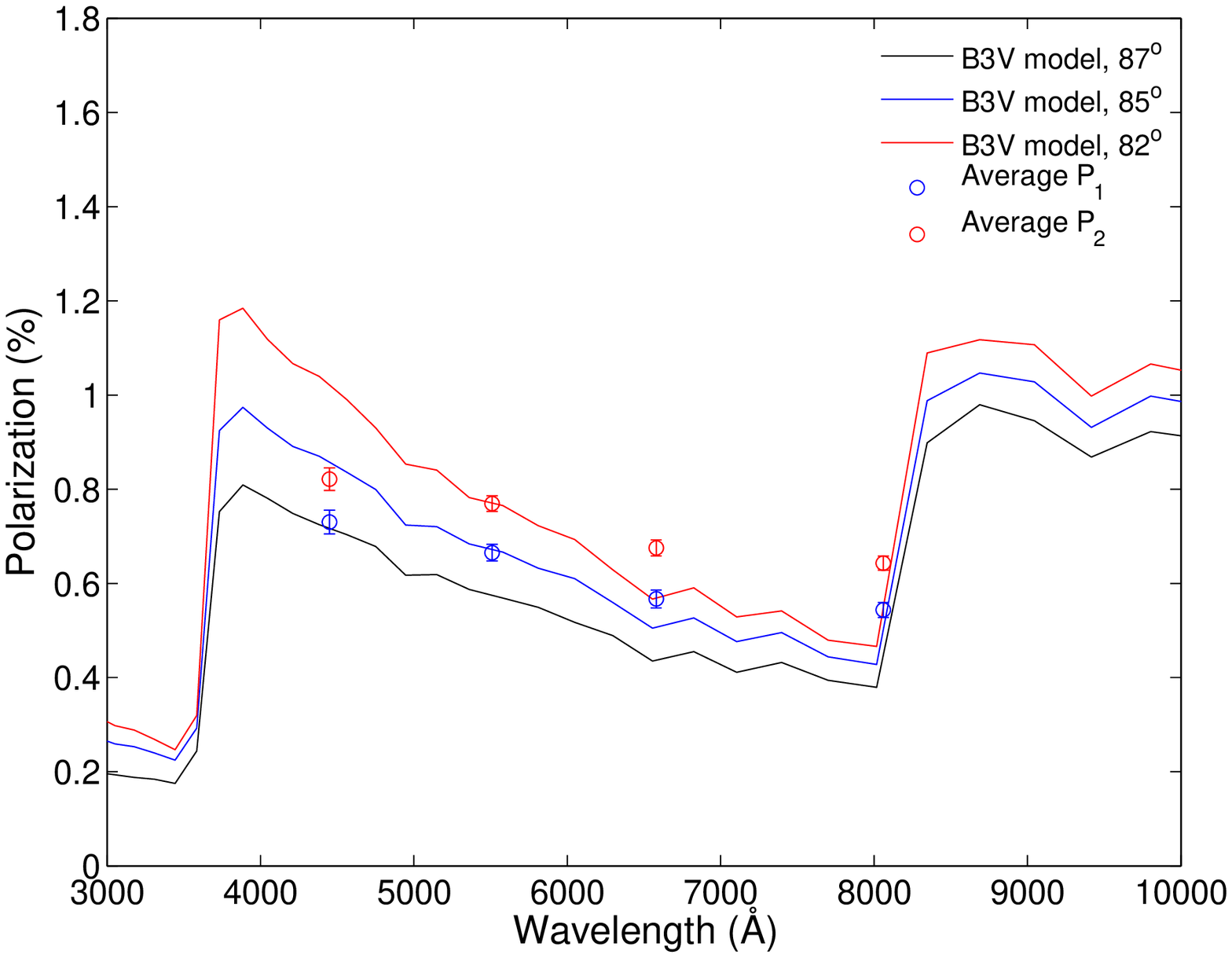}
\caption[Polarization of 48~Librae]{Polarization of 48~Librae. Left
  panel: continuum linear polarization measurements shown with models
  created at different inclination angles, as indicated in the legend.
  Blue dots ($P_{\rm{1}}$) represent the observations corrected by the
  ISP given in \citet{ste12}, and red dots ($P_{\rm{2}}$) represent
  the observations corrected by the ISP given in \citet{dra14}.  Right
  panel: Averaged $P_{\rm{1}}$ and $P_{\rm{2}}$ continuum linear
  polarization measurements with their averaged errors, bracketed by
  models at 82\degr and 87\degr.  The best average representation is
  given by the model created at 85\degr. \label{fig:48libpol}}
\end{figure*}

\begin{figure}
\centering
\includegraphics[scale=0.6]{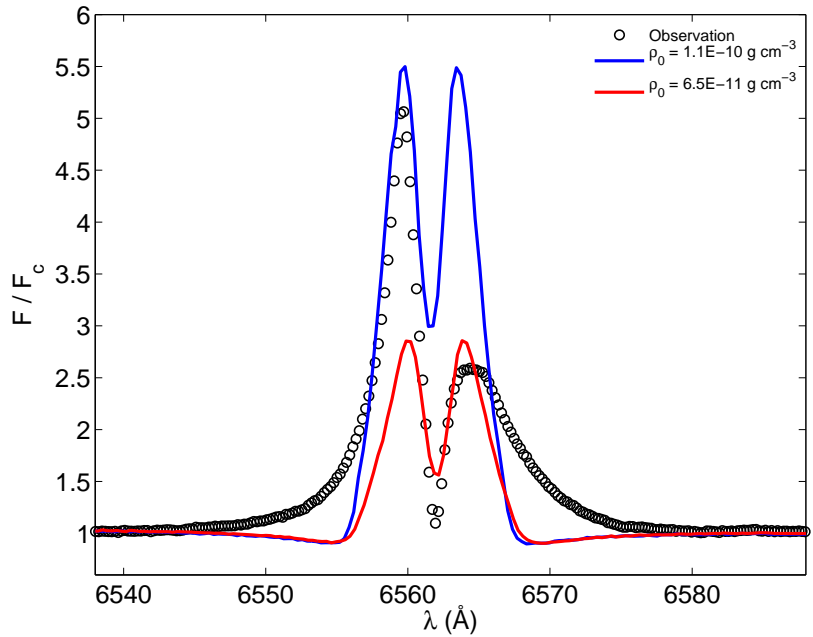}
\caption[H$\alpha$ profile of 48~Librae]{H$\alpha$ profile of
  48~Librae.  The observed profile, shown in open circles, has each
  of its peaks matched individually by unperturbed disk models of
  initial density as specified in the legend. \label{fig:ha48}}
\end{figure}

\begin{figure}
\centering
\includegraphics[scale=0.6]{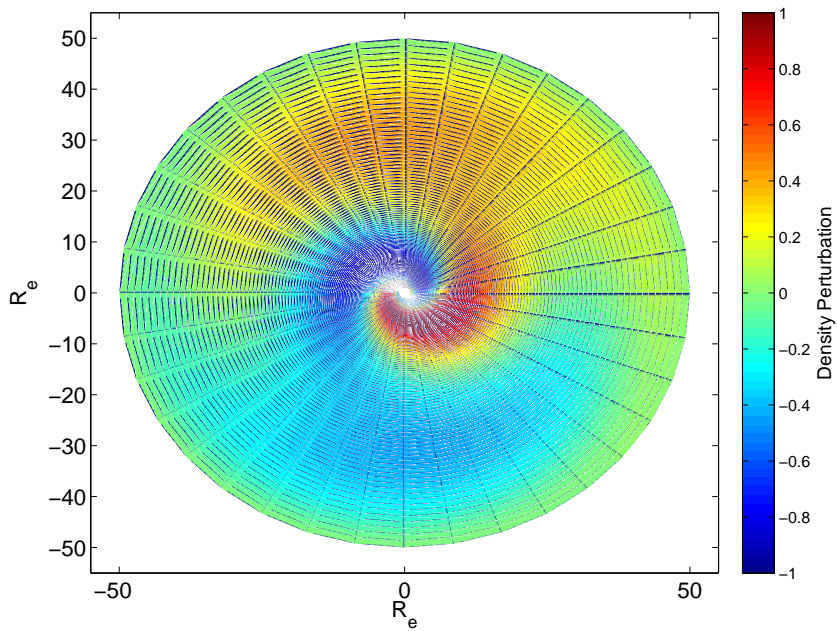}
\caption{Density perturbation pattern as predicted by the 2D global
  oscillation code with $\alpha$ = 0.76.  The disk is viewed from
  above, with a void at the centre representing the location of the
  central star, and the $x$- and $y$-axes give the distance from the
  central star in units of stellar radii.  The density perturbation is
  normalized by the unperturbed value, with red regions corresponding
  to overdense regions and blue regions corresponding to underdense
  regions. \label{fig:pert}}
\end{figure}

\begin{figure*}
\includegraphics[scale=0.8]{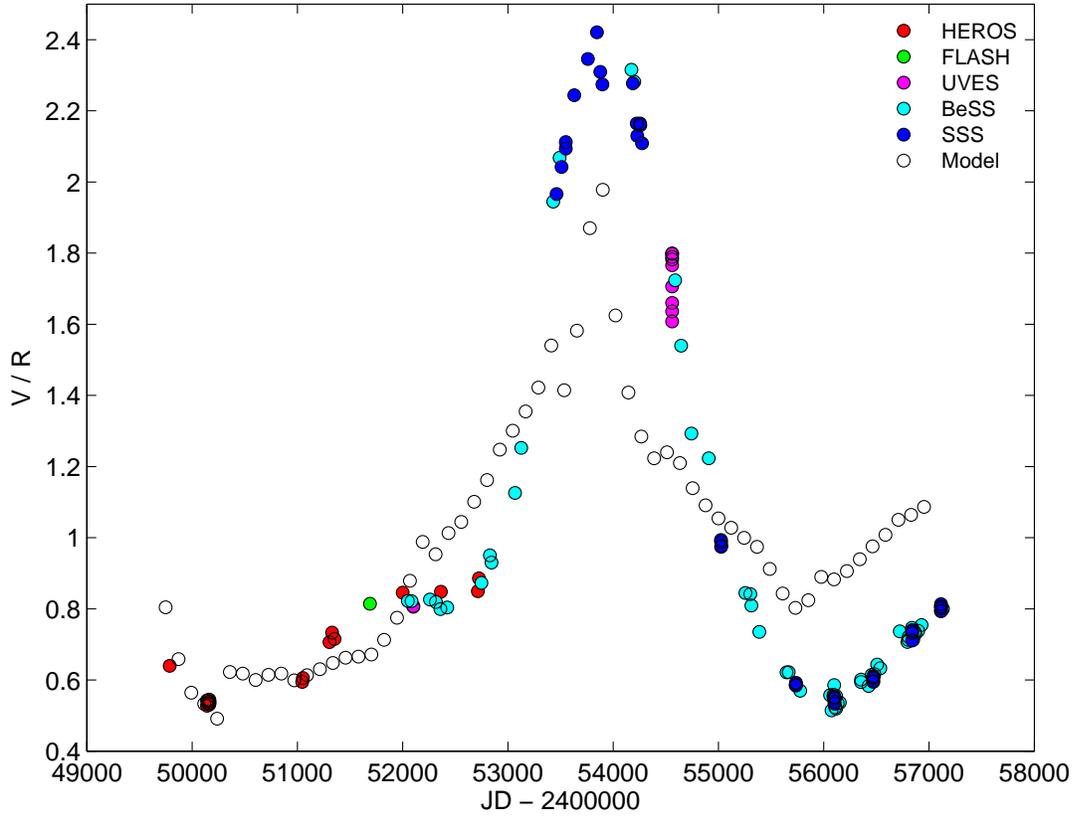}
\caption{$V/R$ cycle of 48~Librae. Filled circles represent $V/R$
  observations, with different colours representing different
  instruments used to collect the H$\alpha$ line spectra, as given in
  the legend. The model H$\alpha$ $V/R$ cycle computed from the
  perturbed disk shown in Figure~\ref{fig:pert} is overplotted in open
  circles. The model cycle has been shifted in phase to align with the
  observed cycle, has been plotted over the same time period as the
  observations.  \label{fig:vrcycle}}
\end{figure*}

\begin{figure}
\centering
\includegraphics[scale=0.8]{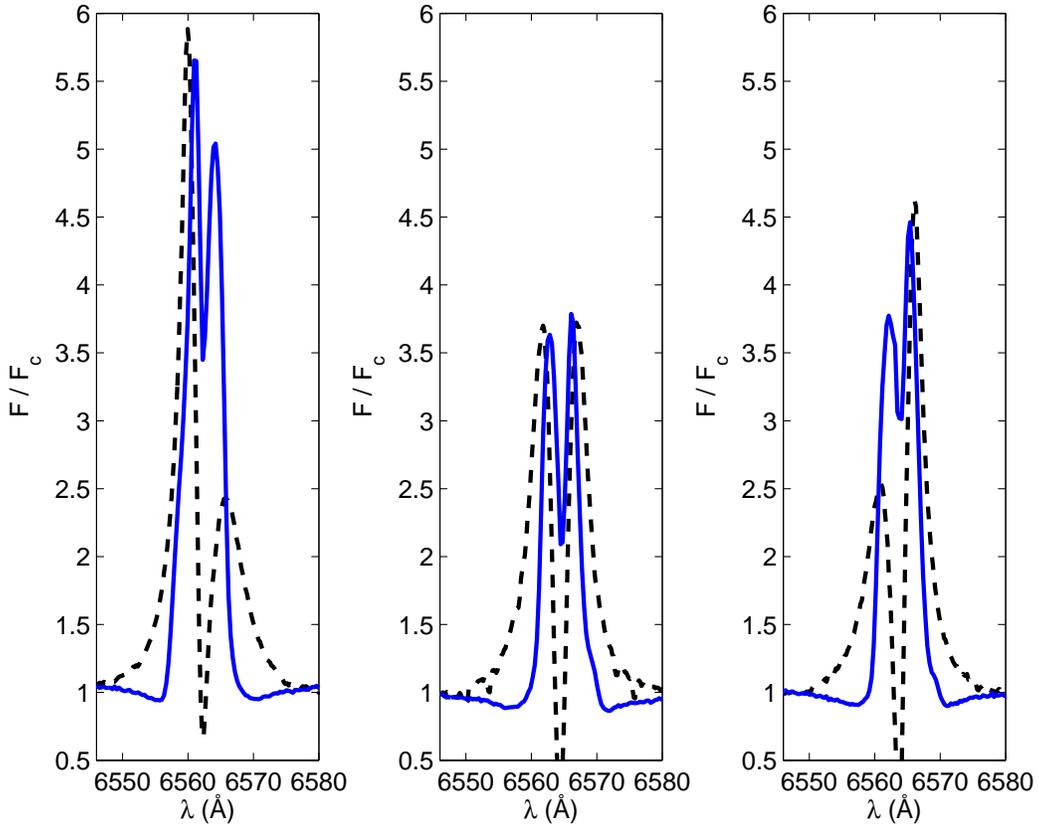}
\caption{Comparison of observed (dashed line) and model (solid blue
  line) H$\alpha$ profiles at three different phases.  Left panel: the
  observed H$\alpha$ profile, obtained on March 26, 2007, is in a $V >
  R$ configuration, and the model is at $\varphi = 252\degr$. Centre
  panel: the observed H$\alpha$ profile, obtained on July 10, 2009, is
  in a $V$ = $R$ configuration, and the model is at $\varphi =
  66\degr$.  Right panel: the observed H$\alpha$ profile, obtained on
  June 22, 2012, is in a $V < R$ configuration, and the model is at
  $\varphi = 36\degr$. \label{fig:phases}}
\end{figure}

\clearpage

\begin{deluxetable}{lrcc}
\tablecolumns{4}
\tablewidth{0pt}
\tablecaption{Photometric Observations \label{tab:phot}}
\tablehead{
\colhead{Instrument} & \colhead{$\lambda$} & \colhead{Flux} & \colhead{Error}\\
\colhead {and/or Filter} & \colhead{(\AA)}  & \colhead{(erg\,s$^{-1}$\,cm$^{-2}$\,\AA$^{-1}$)} & \colhead{(erg\,s$^{-1}$\,cm$^{-2}$\,\AA$^{-1}$)} 
}

\startdata
IUE 1250-1300        & 1285            & 4.583E-10      & 2.529E-11 \\
IUE 1450-1500        & 1475            & 4.800E-10      & 2.451E-11 \\ 
IUE 1675-1725        & 1700            & 3.226E-10      & 1.275E-11  \\
IUE 2150-2200        & 2175            & 1.593E-10      & 1.389E-11 \\
IUE 2395-2445        & 2420            & 1.343E-10      & 1.111E-11 \\
IUE 2900-3000        & 2950            & 1.015E-10      & 4.856E-12 \\
Str\"{o}mgren u      & 3448            & 1.747E-11      & 4.033E-11 \\
Johnson U            & 3571            & 5.475E-11      & 2.171E-12 \\
Str\"{o}mgren v      & 4110            & 7.771E-11      & 3.860E-11  \\
TYCHO B              & 4280            & 7.695E-11      & 9.922E-13 \\
Johnson B            & 4378            & 7.678E-11      & 1.773E-12 \\
Str\"{o}mgren b      & 4663            & 6.473E-11      & 7.003E-12 \\
TYCHO V              & 5340            & 4.260E-11      & 3.531E-13 \\
Johnson V            & 5466            & 4.021E-11      & 8.519E-13 \\
Str\"{o}mgren y      & 5472            & 3.976E-11      & 4.321E-12 \\
2MASS J              & 12350           & 2.858E-12      & 5.687E-13 \\
2MASS H              & 16620           & 1.326E-12      & 9.772E-14 \\
2MASS Ks             & 21590           & 6.242E-13      & 1.149E-14 \\
WISE W1              & 33526           & 1.484E-13      & 1.038E-14 \\
WISE W2              & 46028           & 8.947E-14      & 4.532E-15 \\
IRAS 12$\mu$m        & 101465          & 4.688E-15      & 1.875E-16 \\
WISE W3              & 115608          & 3.176E-15      & 3.510E-17 \\
AKARI IRC L18W       & 176095          & 1.053E-15      & 1.430E-17 \\
IRAS 25$\mu$m        & 217265          & 3.772E-16      & 4.149E-17 \\
WISE W4              & 220883          & 5.349E-16      & 1.083E-17 \\
IRAS 60$\mu$m        & 519887          & 4.436E-17\tablenotemark{a}      & \nodata   \\
IRAS 100$\mu$m       & 952971          & 6.800E-17\tablenotemark{a}      & \nodata   \\
APEX                 & 8700000         & 7.306E-21      & 1.796E-21 \\
\enddata
\tablenotetext{a}{Value is considered to be an upper limit only.}
\end{deluxetable}

\begin{deluxetable}{llccrr}
\tablewidth{0pt}
\tablecolumns{6}
\tablecaption{H$\alpha$ Spectroscopic Observations \label{tab:Ha}}
\tablehead{
\colhead{Telescope} & \colhead{Instrument} & \colhead{Date} & \colhead{JD} & \colhead{No. of} & \colhead{Resolving}\\
\colhead{Name} & \colhead{Name} & \colhead{} & \colhead{2400000+} & \colhead{Spectra} & \colhead{Power}
}

\startdata
ESO 50 cm, Ond\v{r}ejov 2 m  & \textsc{heros} & 1995--2003 & 49\,788--52\,725 & 18 & 20\,000 \\
Wendelstein 80 cm            & \textsc{flash} & 2000-05-25 & 51\,690          & 1  & 20\,000 \\
ESO/VLT-Kueyen               & \textsc{uves}  & 2001-07-11 & 52\,101          & 2  & 60\,000 \\
ESO/VLT-Kueyen               & \textsc{uves}  & 2008-04-03 & 54\,559          & 9  & 60\,000 \\    
John S. Hall 1.1 m           & \textsc{sss}   & 2005--2015 & 53\,462--57\,114 & 49 & 10\,000 \\

\cline{1-6} \\
\multicolumn{6}{c}{BeSS Database} \\
\cline{1-6}
CN212, FS128     & \textsc{lhires1}    & 2001--2004  & 52\,048--53\,126 & 11  & 6\,000   \\
C8, C11, C12     & \textsc{lhires3}    & 2005--2014  & 53\,430--56\,802 & 19  & 17\,000  \\
CNC212           & Echelle V1          & 2008-05-01  & 54\,588          & 1   & 8\,000   \\ 
SC12             & \textsc{lhires-a12}t& 2008        & 54\,645, 54\,743 & 2   & 6\,000   \\
C11              & eShel               & 2009--2010  & 54\,907--55\,301 & 3   & 10\,000  \\
CNC212, C11, C14 & \textsc{lhires3}    & 2010--2015  & 55\,388--57\,131 & 16  & 15\,000  \\
C9               & \textsc{lhires3}    & 2012-05-12  & 56\,060          & 1   & 12\,000  \\ 
C14              & \textsc{lhires3}    & 2012-06-20  & 56\,098          & 1   & 11\,000  \\
\enddata
\end{deluxetable}

\clearpage

\begin{deluxetable}{cccccc}
\tablewidth{0pt}
\tablecolumns{6}
\tablecaption{Polarimetric Observations \label{tab:pol}}
\tablehead{
\colhead{Obs.} & \colhead{MJD} & \colhead{Filter} & \colhead{$\lambda$ (\AA)} & \colhead{$P_{\rm{1}}$\tablenotemark{a} (\%)} & \colhead{$P_{\rm{2}}$\tablenotemark{b} (\%)} \\
}

\startdata
1  & 54975.649 &	B &	4450 &	0.777 $\pm$ 0.019 & 0.868 $\pm$ 0.019 \\
   & 54975.660 &	V &	5510 &	0.733 $\pm$ 0.020 & 0.830 $\pm$ 0.020 \\
   & 54975.682 &	R &	6580 &	0.603 $\pm$ 0.021 & 0.699 $\pm$ 0.021 \\ 
2  & 55012.623 &	B &	4450 &	0.762 $\pm$ 0.013 & 0.866 $\pm$ 0.008 \\
   & 55012.578 &	V &	5510 &	0.714 $\pm$ 0.013 & 0.847 $\pm$ 0.006 \\
   & 55012.601 &	R &	6580 &	0.646 $\pm$ 0.009 & 0.742 $\pm$ 0.009 \\
3  & 55334.752  &	B &	4450 &	0.827 $\pm$ 0.029 & 0.888 $\pm$ 0.029 \\
   & 55334.761  &	V &	5510 &	0.699 $\pm$ 0.025 & 0.796 $\pm$ 0.023 \\
   & 55334.772  &	R &	6580 &	0.559 $\pm$ 0.032 & 0.615 $\pm$ 0.016 \\
   & 55334.764  &	I &	8060 &	0.579 $\pm$ 0.035 & 0.738 $\pm$ 0.028 \\
4  & 55396.521  &	B & 	4450 &	0.754 $\pm$ 0.017 & 0.849 $\pm$ 0.017 \\
   & 55396.515  &	V &	5510 &	0.700 $\pm$ 0.018 & 0.810 $\pm$ 0.021 \\
   & 55396.503  &   	R &	6580 &	0.519 $\pm$ 0.017 & 0.632 $\pm$ 0.011 \\
   & 55396.492  &	I &	8060 &	0.546 $\pm$ 0.014 & 0.644 $\pm$ 0.007 \\
5  & 55427.399  &	B &	4450 &	0.710 $\pm$ 0.017 & 0.833 $\pm$ 0.017 \\
   & 55427.415  &	V &	5510 &	0.632 $\pm$ 0.023 & 0.744 $\pm$ 0.023 \\
   & 55427.428  &	R &	6580 &	0.408 $\pm$ 0.025 & 0.692 $\pm$ 0.025 \\
   & 55427.437  &	I &	8060 &	0.524 $\pm$ 0.020 & 0.617 $\pm$ 0.020 \\
6  & 55441.388  &	B &	4450 &	0.756 $\pm$ 0.018 & 0.791 $\pm$ 0.009 \\
   & 55441.396  &	V &	5510 &	0.696 $\pm$ 0.022 & 0.787 $\pm$ 0.022 \\
   & 55441.403  &	R &	6580 &	0.594 $\pm$ 0.020 & 0.693 $\pm$ 0.018 \\
   & 55441.409  &	I &	8060 &	0.522 $\pm$ 0.022 & 0.616 $\pm$ 0.024 \\
7  & 55695.701  &	B &	4450 &	0.608 $\pm$ 0.022 & 0.739 $\pm$ 0.017 \\
   & 55695.686  &	V &	5510 &	0.619 $\pm$ 0.013 & 0.741 $\pm$ 0.004 \\
   & 55695.696  &	R &	6580 &  0.563 $\pm$ 0.005 & 0.666 $\pm$ 0.006 \\
8  & 56408.085  &  	B &	4450 &	0.819 $\pm$ 0.150 & 0.909 $\pm$ 0.150 \\
   & 56408.102  &	V &	5510 & 	0.641 $\pm$ 0.038 & 0.738 $\pm$ 0.038 \\
   & 56408.118  &       R & 	6580 &	0.589 $\pm$ 0.041 & 0.685 $\pm$ 0.041 \\
   & 56408.137  &	I &	8060 & 	0.530 $\pm$ 0.017 & 0.624 $\pm$ 0.017 \\
9  & 56408.245  &	B &	4450 &	0.661 $\pm$ 0.007 & 0.750 $\pm$ 0.007 \\
   & 56408.252  &  	V &	5510 &	0.568 $\pm$ 0.018 & 0.664 $\pm$ 0.018 \\
   & 56408.262  &	R &	6580 &	0.474 $\pm$ 0.016 & 0.570 $\pm$ 0.016 \\
   & 56408.270  &	I &	8060 &	0.488 $\pm$ 0.020 & 0.580 $\pm$ 0.020 \\
10 & 56505.053  &	B &	4450 &	0.736 $\pm$ 0.009 & 0.826 $\pm$ 0.009 \\
   & 56505.060  &	V &	5510 &	0.654 $\pm$ 0.005 & 0.751 $\pm$ 0.005 \\
   & 56505.070  & 	R &	6580 &	0.599 $\pm$ 0.008 & 0.695 $\pm$ 0.008 \\
   & 56505.080  &	I &	8060 &	0.554 $\pm$ 0.004 & 0.646 $\pm$ 0.004 \\
11 & 56517.025  &	B &	4450 &	0.676 $\pm$ 0.011 & 0.766 $\pm$ 0.011 \\
   & 56517.038  &  	V &	5510 &	0.608 $\pm$ 0.009 & 0.704 $\pm$ 0.009 \\
   & 56517.048  &	R &	6580 &	0.511 $\pm$ 0.011 & 0.607 $\pm$ 0.011 \\
   & 56517.057  &	I &	8060 &	0.507 $\pm$ 0.010 & 0.601 $\pm$ 0.010 \\
12 & 56839.992  & 	B &	4450 &	0.746 $\pm$ 0.017 & 0.836 $\pm$ 0.017 \\
   & 56839.996  &	V &	5510 &	0.722 $\pm$ 0.025 & 0.817 $\pm$ 0.025 \\
   & 56840.001  &	R &	6580 &	0.661 $\pm$ 0.024 & 0.758 $\pm$ 0.024 \\
   & 56840.008  &	I &	8060 &	0.578 $\pm$ 0.012 & 0.671 $\pm$ 0.012 \\
13 & 56888.979  &	B &	4450 &	0.705 $\pm$ 0.006 & 0.796 $\pm$ 0.006 \\
   & 56888.984  &	V &	5510 &	0.645 $\pm$ 0.008 & 0.759 $\pm$ 0.008 \\
   & 56888.990  &	R &	6580 &	0.581 $\pm$ 0.021 & 0.668 $\pm$ 0.021 \\
   & 56888.998  &	I &	8060 &	0.549 $\pm$ 0.009 & 0.633 $\pm$ 0.009 \\
14 & 57142.114  &	B &	4450 &	0.688 $\pm$ 0.016 & 0.777 $\pm$ 0.016 \\
   & 57142.103  &	V &	5510 &	0.683 $\pm$ 0.012 & 0.779 $\pm$ 0.012 \\
   & 57142.108  &	R &	6580 &  0.631 $\pm$ 0.016 & 0.727 $\pm$ 0.016 \\
   & 57142.120  &	I &	8060 &	0.603 $\pm$ 0.012 & 0.696 $\pm$ 0.012 \\
\enddata
\tablenotetext{a}{Polarization value calculated using the ISP given in \citet{ste12}.}
\tablenotetext{b}{Polarization value calculated using the ISP given in \citet{dra14}.}
\end{deluxetable}

\clearpage

\begin{deluxetable}{llc} 
\tabletypesize{\scriptsize} 
\tablecolumns{3} 
\tablewidth{0pt} 
\tablecaption{Adopted Stellar and Disk Parameters of 48~Librae \label{tab:48params}} 
\tablehead{ 
\colhead{Parameter} & \colhead{Value} & \colhead{Ref.} \\
} 
\startdata 
\multicolumn{3}{c}{Stellar Parameters} \\ 
Spectral Type   & B3V            & 1,2   \\ 
$M_\star$        & 6.07\,$M_\odot$   & 1,3 \\
$R_{\rm{p}}$      & 3.12\,$R_\odot$   & 4   \\
$R_{\star}$       & 4.12\,$R_\odot$       & 1  \\    
$L_\star$        & 1100\,$L_\odot$   & 4   \\
$W$             & 0.74            & 1    \\
$\beta_{GD}$     & 0.17           & 5     \\
\cline{1-3} \\
\multicolumn{3}{c}{Disk Parameters} \\ 
$\rho_{0,V}$   & 1.1 $\times10^{-10}$ g\,cm$^{-3}$ & 1  \\ 
$\rho_{0,R}$   & 6.5 $\times10^{-11}$ g\,cm$^{-3}$ & 1  \\ 
$\rho_{0,avg}$  & 8.8 $\times10^{-11}$ g\,cm$^{-3}$  & 1  \\
\cline{1-3} \\
\multicolumn{3}{c}{$V/R$ Parameters} \\
$P$           & 12.13 yr            &   1        \\      
$k_2$         & 0.006                &  6         \\
$\eta$        & 5.74 $\times10^{-2}$  &  7         \\ 
$\epsilon$    & 0.1                  &  7         \\
$\alpha$      & 0.76                 &  1          \\
$\dot{M}$     & 8.27 $\times10^{-10}M_\odot$\,yr$^{-1}$ &   1         \\
\cline{1-3}\\
\multicolumn{3}{c}{Geometrical Parameters} \\ 
$i$   & $85 \pm 3$\degr  & 1  \\ 
$d$   & $143\substack{+5 \\ -10}$ pc    & 8  \\ 
\enddata \tablerefs{ (1)~this work; (2)~\citet{und53};
  (3)~\citet{har88}; (4)~\citet{geo13}; (5)~\citet{esp12};
  (6)~\citet{pap92}; (7)~\citet{car09}; (8)~\citet{van07}.  } 
\end{deluxetable} 

\clearpage

\bibliography{refs}
\end{document}